\documentclass[twocolumn,superscriptaddress,showpacs,prl]{revtex4}
\usepackage{epsfig}
\usepackage{delarray}
\usepackage{amsmath}
\usepackage{bm}

\newcommand{\ket}[1]{| \, #1 \rangle}
\newcommand{\bra}[1]{ \langle #1 \,  |}

\begin{document}
\title{Local transformation of two EPR photon pairs into a three-photon W state}

\author{Toshiyuki Tashima}
%\email{tashima@qi.mp.es.osaka-u.ac.jp} 
\affiliation{Graduate School of Engineering Science, Osaka University, Toyonaka, Osaka 560-8531, Japan}
\affiliation{CREST Photonic Quantum Information Project 4-1-8 Honmachi,
Kawaguchi, Saitama 331-0012, Japan}
\author{Tetsuroh Wakatsuki}
\affiliation{Graduate School of Engineering Science, Osaka University, Toyonaka, Osaka 560-8531, Japan}
\affiliation{CREST Photonic Quantum Information Project 4-1-8 Honmachi, Kawaguchi, Saitama 331-0012, Japan}
\author{\c{ S}ahin Kaya \"Ozdemir}
%\email{ozdemir@qi.mp.es.osaka-u.ac.jp}
\affiliation{Graduate School of Engineering Science, Osaka University, Toyonaka, Osaka 560-8531, Japan}
\affiliation{CREST Photonic Quantum Information Project 4-1-8 Honmachi, Kawaguchi, Saitama 331-0012, Japan}
\affiliation{ERATO Nuclear Spin Electronics Project, Aramaki, Aza Aoba, Sendai 980-0845, Japan}
\author{Takashi Yamamoto}
%\email{}
\affiliation{Graduate School of Engineering Science, Osaka University, Toyonaka, Osaka 560-8531, Japan}
\affiliation{CREST Photonic Quantum Information Project 4-1-8 Honmachi, Kawaguchi, Saitama 331-0012, Japan}
\author{Masato Koashi}
%\email{}
\affiliation{Graduate School of Engineering Science, Osaka University, Toyonaka, Osaka 560-8531, Japan}
\affiliation{CREST Photonic Quantum Information Project 4-1-8 Honmachi, Kawaguchi, Saitama 331-0012, Japan}
\author{Nobuyuki Imoto}
%\email{}
\affiliation{Graduate School of Engineering Science, Osaka University, Toyonaka, Osaka 560-8531, Japan}
\affiliation{CREST Photonic Quantum Information Project 4-1-8 Honmachi, Kawaguchi, Saitama 331-0012, Japan}

%\date{\today}
\begin{abstract}
We propose and experimentally demonstrate a transformation of two EPR
 photon pairs distributed among three parties into a three-photon W
 state using local operations and classical communication. We then characterize the final state using quantum state tomography on
the three-photon state and on its marginal bipartite states. The fidelity of the
final state to the ideal W state is $0.778\pm 0.043$ and the
expectation value for its witness operator is $-0.111\pm 0.043$
implying the success of the proposed local transformation.
\pacs{03.67.Hk, 03.65.Ud}
\end{abstract}
\maketitle
%%%%%%%%%%%%%%%%%%%%%%%%%%%%%%% main %%%%%%%%%%%%%%%%%%%%%%%%%%%%%

Recent development in quantum
information science has revealed most of the mysteries concerning entanglement between two qubits:
We know how to prepare, characterize and
quantify it. In particular, 
using local operations and classical communication
(LOCC), any state of two qubits can be generated 
from a single resource of qubits in
a maximally entangled state, which is called 
an Einstein-Podolsky-Rosen (EPR) pair. For photonic qubits, the state
$\ket{{\rm EPR}}\equiv (\ket{\rm HH}+\ket{\rm VV})/\sqrt{2}$
serves as an EPR pair, 
where $\ket {\rm H}$ and $\ket {\rm V}$
represents the horizontal and the vertical polarization,
respectively. By contrast, entanglement among three or more qubits
still remains as
a challenge because such systems have a richer and more complex
structure which originates from the existence of different ways the
qubits can be entangled with each other. One of such
interesting features is the presence of inequivalent classes of
entangled states: The states in distinct classes cannot be
interconverted using LOCC \cite{s1}, even probabilistically. 
The simplest example is 
the three-qubit entanglement, where 
there are two distinct classes of states: 
the Greenberger-Horne-Zeilinger (GHZ) type of 
states including a 
standard one $\ket {\rm GHZ_3} = (\ket {\rm HHH}+\ket {\rm VVV})/\sqrt
2$,
 and the W-type states including
$\ket {\rm W_3}=(\ket {\rm HHV}+\ket {\rm HVH}+\ket {\rm
VHH})/\sqrt 3$. A key distinction between these states is that while
the loss of any one of the qubits completely destroys the
entanglement 
in the GHZ state $\ket {\rm GHZ_3}$,
the entanglement between the
remaining two qubits survives in
the W state $\ket {\rm W_3}$ \cite{s1}.
Recently, there have been a number of theoretical
proposals and experimental demonstrations for the preparation of
three-qubit GHZ and W states in optical experiments \cite{s2,s3,s4.0,s4,s5,s6,s7,s8,s9,s10}.

Existence of the distinct classes implies that 
there is no three-qubit state that can be
used as a universal resource for generating 
arbitrary three-qubit pure states under LOCC. For this purpose,
one must look for a resource in larger systems.
One of the simplest way is to distribute the resource 
for bipartite entanglement between one party 
(Charlie) and each of the other parties (Alice and Bob),
resulting in state $\ket{{\rm EPR}}_{\rm AC}\ket{{\rm EPR}}_{\rm BC'}$.
Starting with this resource, it is at least theoretically easy to show
that Charlie can prepare three local auxiliary qubits in the desired
three-qubit state, which may be an entangled state, and then faithfully send one qubit to Alice and another to Bob by quantum teleportation \cite{s11}. Since this scenario involves seven qubits
in total, it is hard to carry out in experiments and a
more direct way of converting EPR pairs to three-qubit states 
is desired. For the GHZ-type states, it is easy to do this since 
we can convert the two EPR pairs to $\ket {\rm GHZ_3}$ by quantum parity 
check \cite{s2,s11.1}, which can be done by a polarization beam splitter and
post-selection\cite{s6}.
 Any GHZ-type state is then produced with nonzero probability 
by applying a unitary operation and local filtering on each photon, which can be 
done with high precision. This line of strategy was further extended for 
the W-type states by Walther {\it et al.} \cite{s7},
who experimentally demonstrated that $\ket {\rm W_3}$ can be 
{\it approximately} generated from $\ket {\rm GHZ_3}$ by LOCC.
In this method, there is a trade-off
between the success probability and the fidelity of the final
state such that the fidelity approaches unity only in the limit of
zero success probability, which reflects the fact that 
$\ket {\rm GHZ_3}$ and $\ket {\rm W_3}$ belong to distinct classes
of states.

The aim of this paper is to propose and experimentally demonstrate the 
missing path of resource conversion, namely, direct transformation 
of bipartite resource $\ket{{\rm EPR}}_{\rm AC}\ket{{\rm EPR}}_{\rm BC'}$ into
the W state $\ket {\rm W_3}$. Our
 scheme simply uses a polarization-dependent beam splitter (PDBS)
and a photon detection to realize a desired transformation of Charlie's 
two photons into one photon. We will first 
discuss the working principle and then describe our experimental
results.

\begin{figure}[tbp]
\begin{center}
\includegraphics[scale=1]{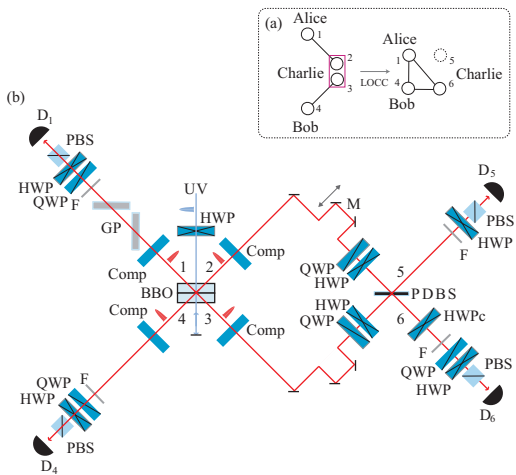}
\caption{Schematics of the experiment. (a) The concept of our conversion
 from two EPR pairs to the W state. EPR pairs are shared by
 Alice-Charlie and Bob-Charlie. Charlie's operation is performed on his
 two qubits. (b) The experimental setup. The
ultraviolet (UV) pulses (wavelength 395nm, average power 380mW, diagonal polarization) from a
frequency-doubled mode-locked Ti:Sapphire laser (wavelength 790nm;
pulse width 90fs; repetition rate 82MHz) make two passes through a
pair of Type-I phase matched $\beta$-barium borate (BBO) crystals
(thickness 1.5mm) stacked with their optical axes orthogonal to
each other to produce two EPR photon pairs 
via spontaneous parametric down conversion (SPDC) \cite{s15,s15.1}. Extra BBOs
with thickness 1.65mm (comp) are placed on the
path of the each photon to compensate for walk-off effects. The spectral
 filtering of the photons is done by a narrow-band interference filter
 (IF, wavelength: 790nm; bandwidth: 2.7nm). All the detectors $\rm D_1$, $\rm D_4$, $\rm D_5$
 and $\rm D_6$ are silicon avalanche photodiodes placed after single-mode
 optical fibers to achieve high fidelity.\label{fig:1s}}
\end{center}
\end{figure}

Let us assume that four photons in state 
$\ket {\rm EPR}_{12}\ket
{{\rm EPR}}_{34}=(\ket{\rm HHHH}_{1234}+\ket{\rm HHVV}_{1234}+\ket{\rm
VVHH}_{1234}+\ket{\rm VVVV}_{1234})/2$ 
are distributed such that 
Alice has the photon in mode 1, Bob has mode 4,
and Charlie has modes 2 and 3.
Charlie sends his two photons to a PDBS,
whose output modes are labelled as 5 and 6.
Let $\mu>0$ be the transmission coefficient of the PDBS
for the H polarization, and $\nu>0$ for the V polarization. We are only interested in the case where 
a photon is present in each of the modes 1, 4, 5 and 6.
Keeping only such terms, the state after the PDBS
is written as 
\begin{eqnarray}
&&[c \ket{{\rm HHH}}_{146}+ b \ket{{\rm HVV}}_{146} + a \ket{{\rm VHV}}_{146}]
\ket {{\rm H}}_{5}
\nonumber\\&&
+[d \ket{{\rm VVV}}_{146}+ a \ket{{\rm HVH}}_{146} + b \ket{{\rm VHH}}_{146}]
\ket {{\rm V}}_{5},
\end{eqnarray}
with
 \begin{eqnarray}
&& a=\sqrt{\mu\nu}/2, \;\;\; b=-\sqrt{(1-\mu)(1-\nu)}/2,
\nonumber\\
&& c=(2\mu-1)/2, \;\;\; d=(2\nu-1)/2.
\end{eqnarray}
If Charlie has detected 
an $\rm H$-polarized photon in mode 5, he announces it and switches 
the polarization of mode 6 as 
$\ket{\rm H}_6\leftrightarrow \ket{\rm V}_6$.
At this point, the three parties share the state
$c \ket{{\rm HHV}}+ b \ket{{\rm HVH}} + a \ket{{\rm VHH}}$.
If the amplitudes $(a, b, c)$ are not equal, they can be 
adjusted by introducing a phase shift and an attenuation for 
V polarization of each photon, resulting in the W state 
$\ket{\rm W_3}$. The overall success probability is given by 
$p_{{\rm H}}\equiv 3 \min\{|a|^2,|b|^2,|c|^2\}$.
Similarly, they can also generate $\ket{\rm W_3}$ for the case
where Charlie has detected a $\rm V$-polarized photon in mode 5,
with an overall success probability 
$p_{{\rm V}}\equiv 3 \min\{|d|^2,|b|^2,|c|^2\}$.
The success probability $p_H$ becomes largest when 
$|a|=|b|=|c|$ holds, which happens when 
the parameters $(\mu, \nu)$ for the PDBS are chosen to be
$\mu=(5+\sqrt 5)/10$ and $\nu=(5-\sqrt 5)/10$
 or vice versa. For this choice, $|a|=|b|=|c|=|d|=1/(2\sqrt{5})$ 
holds, and hence both probabilities take their optimal values
$p_H=p_V=3/20=15\%$ without introducing local attenuations \cite{s14}.

In our experiment, we recorded only the case when 
Charlie has detected an $\rm H$-polarized photon in mode 5.
We also made a sub-optimal choice of the PDBS parameters,
$\mu = (7+\sqrt{17})/16$ and $\nu=1/2$. One of the reasons
for this choice is that the two-photon interference for the 
V polarization is observed directly, which makes the alignment
easier and gives us a clue about how well the 
two photons from different pairs are overlapped at the PDBS.    
Under this choice, we have $|a|>|b|=|c|$ and we need to 
introduce a polarization-dependent loss for Alice's photon
in mode 1. 
The success probability for the ideal case is calculated to be
$p_H=3(9-\sqrt{17})/128\sim 11.4\%$
\begin{figure}[b]
\begin{center}
\includegraphics[scale=1]{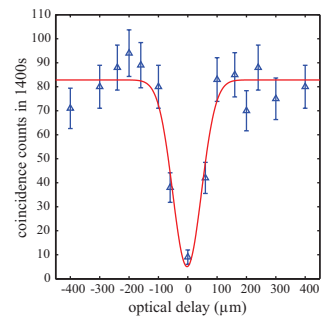}
\caption{Observed two photon interference by recording four-photon
 coincidences as a function of the optical delay. PDBS works as a 50:50
 BS for $\rm V$-polarized photons. The best fit to the
 data is represented by the solid Gaussian curve which shows a coherence length of $l_c \simeq 110 \rm \mu m$. The visibility is $0.885$.\label{fig:2s}}
\end{center}
\end{figure}

The details of our experimental setup are shown in Fig.~\ref{fig:1s}. Charlie's local operations are performed as follows.
Modes 2 and 3 are overlapped at the PDBS, and polarizing beamsplitter (PBS) placed at the output mode 5 selects only the 
H-polarized photons. A half-wave plate (HWPc) at mode 6 interchanges $\rm H$ and $\rm V$
polarizations. On Alice's side,  
a set of glass plates (GP) are placed in mode 1, which can be
tilted to adjust the amount of 
the polarization dependent loss. 
The two plates are tilted in opposite directions such that 
the beam passing through experiences a minimal transverse shift. 
Successful events are signalled by four-photon coincidences using photon
detectors in modes 1, 4, 5 and 6. The
quarter-wave plates (QWP), HWPs and PBSs
in front of the detectors in modes 1, 4 and 6 are used for
verification experiments. 
In order to demonstrate the
effectiveness of the process, we performed experiments to
determine that (i) the photons from the two EPR pairs overlapped well
at the PDBS, (ii) two highly entangled photon pairs
$\rho_{12}$ and $\rho_{34}$  were generated by SPDC, and (iii) the
final three-photon state $\rho_{146}$ was close to the W state $\ket {\rm W_3}$.
\begin{figure}[b]
\begin{center}
\includegraphics[scale=1]{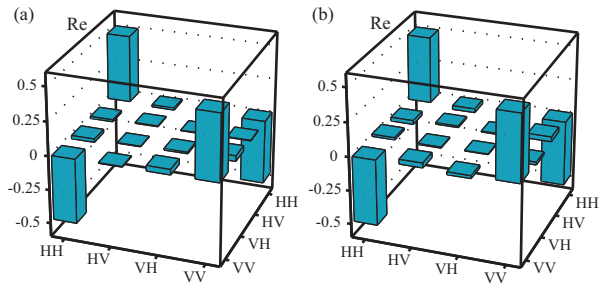}
\caption{Real part of the reconstructed density matrices for the initial
 EPR pairs: (a) $\rho_{12}$ and (b) $\rho_{34}$.\label{fig:3s}}
\end{center}
\end{figure}
\begin{figure}[h]
\begin{center}
\includegraphics[scale=1]{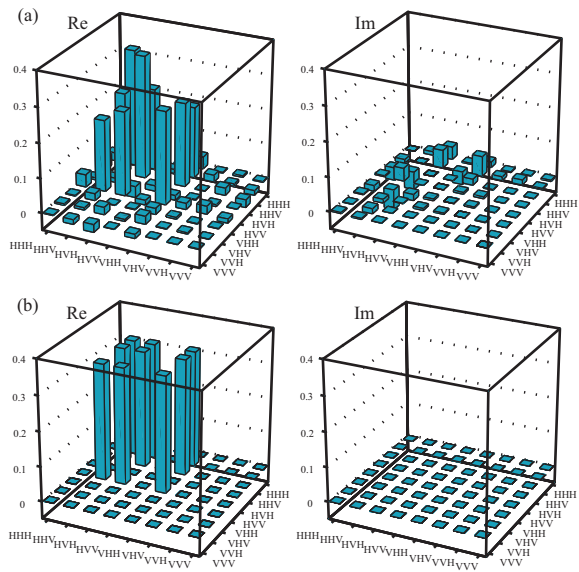}
\caption{(a) Real and imaginary parts of the reconstructed density
 matrix of the experimentally obtained W state, and (b) that
 of the density matrix of the ideal W state $\ket {\rm W_3}$.\label{fig:4s}}
\end{center}
\end{figure}

In order to demonstrate (i), we set the UV pulses to
H-polarization so that in each pass of the UV pulses through the
BBOs, V-polarized photon pairs were generated. The HWPs inserted in front of the detectors were adjusted so that only
V-photons arrive at the detectors. One photon from each pair was
then sent to the PDBS and four-fold coincidences were recorded
while the optical delay experienced by the photons in modes 2 and
3 were changed using the motorized stage M. When the temporal
overlap of these two V-photons at the PDBS was achieved,
Hong-Ou-Mandel dip was observed as in Fig. \ref{fig:2s}. The
observed visibility is $0.885$ at zero delay time.

For (ii), after setting the zero-delay time, we adjusted the UV pulses to
diagonal polarization so that EPR pairs $\rho_{12}$ and
$\rho_{34}$ were generated. Each pair was characterized by 
quantum state tomography (QST)
using 16 different tomographic settings chosen from the
combinations of the single photon projections, $\ket {\rm H}$, $\ket
{\rm V}$, $\ket {\rm D}=(\ket {\rm H}+\ket {\rm V})/\sqrt 2$,
$\ket {{\rm R}}= (\ket {{\rm H}}-i\ket {\rm V})/\sqrt 2$ and $\ket
{{\rm L}}= (\ket {{\rm H}}+i\ket {\rm V})/\sqrt 2$, on each photon \cite{s16}.
Since the PDBS has different transmission coefficients for H and V
polarizations, the measurement bases for the photons in modes 2
and 3 were selected by the HWPs and QWPs inserted before the PDBS. The QWP and HWPs in mode 6 and
the set of glass plates in mode 1 were adjusted so that they did
not affect the polarization of the incoming
photons. Coincidences were recorded in modes 1 and 6 for
$\rho_{12}$, and in modes 4 and 6 for $\rho_{34}$. From these
measured polarization correlations, we estimated the fidelity
$F_{ij}\equiv\bra{{\rm EPR}}\rho_{ij}\ket{{\rm EPR}}$ of each pair to the
ideal EPR pair as $F_{12}=0.967\pm0.002$ and
$F_{34}=0.976\pm 0.002$. 
Here and henceforth, uncertainties in the fidelities and the other
quantities 
were calculated using a Monte Carlo routine assuming Poissonian
statistics of errors.
We further reconstructed their density matrices $\rho_{12}$ and $\rho_{34}$, and calculated the amounts
of entanglement using entanglement of formation (EOF) \cite{s17} as
$0.922\pm0.006$ and $0.947\pm0.004$. The density matrices
estimated using the maximum likelihood method are
shown in Fig. \ref{fig:3s}.

In the last phase of the experiment (iii), we adjusted the
glass plates (GP) to induce the required loss on V-photons in
mode 1,
 and set the QWPs and HWPs in modes 2 and 3 such that they only add a constant phase shift between H and V on the
 incoming photons. HWP in mode 5 was also adjusted so that only H-photons
 arrive at the detector. HWPc in mode 6 was set to swap H and V polarizations.
 We post-selected the successful events with
four-fold coincidences. The final three-photon state $\rho_{146}$
was characterized using 64 different tomographic settings
\cite{s16} implemented by the sets of QWP, HWP and PBS in front of
the detectors in modes 1, 4 and 6. We recorded coincidences for an
acquisition time of 5800s at each tomographic setting. From the recorded
correlations, we reconstructed the density matrix of $\rho_{146}$ using
iterative maximum likelihood (IML) method \cite{s18,s19}. This is
shown in Fig. \ref{fig:4s} together with the density matrix for the ideal
$\ket{\rm W_3}$. The density matrix for the ideal W state consists
of only nine real nonzero terms, namely,
 the diagonal terms corresponding to $\ket
{\rm HHV}$, $\ket {\rm HVH}$ and $\ket {\rm VHH}$ and six off-diagonal
elements corresponding to coherences among these terms. 
It is seen that the
density matrix of the state prepared in our experiment has a similar
structure with nine dominant elements. 
%Some off-diagonal elements and the imaginary components observed in the
%density matrix of the experimentally obtained state can be attributed to
%mode-mismatch and multi-photon effects(?)

Furthermore, from the reconstructed density matrix, 
we calculated the fidelity as
$F\equiv\bra{\rm W_3}\rho_{146}\ket{\rm W_3}=0.778\pm 0.043$. We also calculated the
entanglement-witness of this state using the operator ${\cal W}_{\rm
W}=\frac{2}{3}\openone - \ket {\rm W_3} \bra {\rm W_3}$ to distinguish
it from separable and bi-separable states \cite{s20}. For an ideal
W state, the expectation value of this operator is $-1/3$. We
find ${\rm Tr}({\cal W}_{\rm W}\rho_{\rm 146}) =
-0.111\pm 0.043$ for the final state in our experiment,
 which confirms that $\rho_{\rm 146}$ has a genuine tripartite
 entanglement.% We believe that the deviation
% from the ideal W state can be attributed to multi-photon effects,
% mode-mismatch and residual ambiguity in phase compensation.

\begin{figure}[b]
\begin{center}
\includegraphics[scale=1]{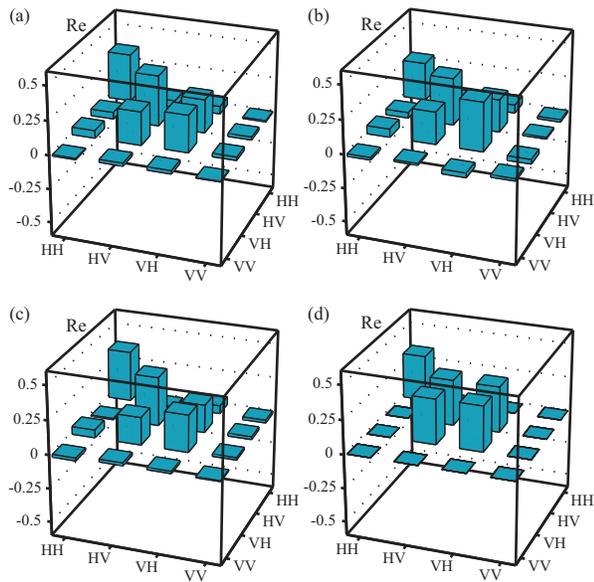}
\caption{Real part of the reconstructed reduced density matrices of the
 experimentally obtained W state, (a) $\rho_{14}$, (b) $\rho_{16}$ and
 (c) $\rho_{46}$. (d) Real part of the reduced density matrix of the
 ideal state $\ket{\rm W_3}$ for which $\rho_{14} = \rho_{16} = \rho_{46}$.\label{fig:5s}}
\end{center}
\end{figure}

One of the distinct properties of the W state is 
the entanglement left in the marginal state of any pair of qubits
after one qubit is removed.
We confirmed this by reconstructing the density
matrices $\rho_{14}$, $\rho_{16}$ and $\rho_{46}$, 
corresponding respectively to Alice-Bob, Alice-Charlie, and Bob-Charlie
marginal bipartite states. 
These density matrices are
given in Fig. \ref{fig:5s} together with the density matrix of the
marginal bipartite state of the ideal W state. 
We also calculated the EOFs as
$0.244\pm0.066$, $0.263\pm0.065$ and $0.195\pm0.065$,
respectively for $\rho_{14}$, $\rho_{16}$ and $\rho_{46}$. It is
clearly seen that Alice and Bob, who initially had no shared
entanglement, now enjoys marginal bipartite entanglement which is created
at the expense of reduction in the initial entanglement between
Alice and Charlie, and the one between Bob and Charlie. 

The imperfection in the final W state produced in our experiment
may be ascribed to the following causes. If we assume that the visibility
observed in  Fig. \ref{fig:2s} reflects the amount of the mode mismatch
between modes 2 and 3, and also assume that it is present independent
of the polarization, this effect is expected to
decrease the final fidelity to 0.89. The imperfections
in the EPR pairs ($F_{12}$ and $F_{34}$) further reduces it to 0.87.
The residual imbalance among the three dominant diagonal terms in Fig. \ref{fig:4s} (a)
explains further reduction by 0.04, arriving in the vicinity of the error bar
of the observed fidelity.

In summary, we have demonstrated 
a method for converting two EPR photon pairs to
a three-photon W state via LOCC,
using a polarization dependent beamsplitter and
post-selection.
The achieved final state was shown to have various characteristics of the W
state. The achieved fidelity of $0.778\pm 0.043$ is higher than the 
value of $0.684\pm0.024$ previously obtained via local transformation
 from a GHZ state \cite{s7}, signifying the advantage of 
direct transformation that does not suffer from 
the fidelity-efficiency trade-off.
This work extends our ability to manipulate multipartite entanglement,
since our results imply that it is now possible to generate arbitrary 
three qubit states from a single resource of two EPR pairs via LOCC
with a moderate succsess probability and with fidelity only limited by 
the imperfection of the apparatus.

This work was partially supported by JSPS Grant-in-Aid for Scientific
Research(C) 20540389 and by MEXT Grant-in-Aid for the Global COE Program and
Young scientists(B) 20740232.


\begin{thebibliography}{99}
\bibitem{s14}
If we restrict ourselves to polarization-independent beamsplitter
	($\mu=\nu$), the optimal choice is $\mu=\nu=2/3$ with $p_H = p_V = 1/12$.
\end{thebibliography}
\end{document}